# Determination of Spin-Orbit Torque Efficiencies in Heterostructures with In-plane Magnetic Anisotropy

Yan-Ting Liu[1], Tian-Yue Chen[1†], Tzu-Hsiang Lo[1], Tsung-Yu Tsai[1], Shan-Yi Yang[3], Yao-Jen Chang[3], Jeng-Hua Wei[3], and Chi-Feng Pai[1,2*]

[1]*Department of Materials Science and Engineering, National Taiwan University, Taipei 10617, Taiwan*

[2]*Center of Atomic Initiative for New Materials, National Taiwan University, Taipei 10617, Taiwan*

[3]*Electronic and Optoelectronic System Research Laboratories, Industrial Technology Research Institute, Hsinchu, Taiwan*

It has been shown that the spin Hall effect from heavy transition metals can generate sufficient spin-orbit torque and further produce current-induced magnetization switching in the adjacent ferromagnetic layer. However, if the ferromagnetic layer has in-plane magnetic anisotropy, probing such switching phenomenon typically relies on tunneling magnetoresistance measurement of nano-sized magnetic tunnel junctions, differential planar Hall voltage measurement, or Kerr imaging approaches. We show that in magnetic heterostructures with spin Hall metals, there exist current-induced in-plane spin Hall effective fields and unidirectional magnetoresistance that will modify their anisotropic magnetoresistance behavior. We also demonstrate that by analyzing the response of anisotropic magnetoresistance under such influences, one can directly and electrically probe magnetization switching driven by the spin-orbit torque, even in micron-sized devices. This pump-probe method allows for efficient and direct determination of key parameters from spin-orbit torque switching events without lengthy device fabrication processes.

---

[†] Email: f05527067@ntu.edu.tw
[*] Email: cfpai@ntu.edu.tw

# I. INTRODUCTION

Spin-orbit torque magnetic random access memory (SOT-MRAM) is a potential candidate architecture to replace now-in-mass-production spin-transfer torque MRAM (STT-MRAM) due to its virtually unlimited endurance, which benefits from the unique SOT writing mechanism that prevents the degradation of tunnel barrier in MRAM cell. The heavy transition metal/ferromagnet (HM/FM) bilayer heterostructures with in-plane magnetic anisotropy (IMA) are commonly adopted for SOT-related research, such as W/CoFeB [1], Ta/CoFeB [2] and Pt/Py [3,4]. The spin current generated from the spin Hall effect (SHE) [5,6] of the HM layer can further exert SOT on the adjacent FM layer. The magnetization in the FM layer then can be switched via an anti-damping mechanism [7]. Unlike the cases where the FM layers have perpendicular magnetic anisotropy (PMA), whose magnetization states can be easily probed by anomalous Hall voltages [8-11], the SOT-induced magnetization switching in HM/FM heterostructures with IMA are typically probed by tunneling magnetoresistance (TMR) measurement on nano-sized three-terminal devices [1,2,12,13], differential planar Hall effect (DPHE) in Hall-bar devices [14], and magneto-optic Kerr effect (MOKE) imaging [15] to further estimate the SOT efficiencies. The challenging device fabrication processes before TMR measurement, the complicated measurement protocol of DPHE (alternating magnetic field is required), and the non-electrical probing of magnetization through MOKE imaging are the major obstacles to develop these characterization methods into an efficient approach to meet industrial needs.

In this work, we present a reliable and simple all-electrical strategy to characterize the SOTs,

both damping-like and field-like, in micron-sized magnetic heterostructures with IMA. Firstly, a current-induced shift of the anisotropic magnetoresistance (AMR) loop and an unidirectional magnetoresistance (UMR) are observed in W(3)/Co$_{40}$Fe$_{40}$B$_{20}$($t_{\text{Co-Fe-B}}$) heterostructures with IMA (numbers in parentheses are in nanometers). The shift in the AMR indicates the existence of a current-induced in-plane effective field, which is originated from the SHE of W. Secondly, the electrical detection of current-induced SOT-driven magnetization switching in a W/Co$_{40}$Fe$_{40}$B$_{20}$ device is demonstrated by means of AMR measurement under the influence of this spin Hall effective field. From switching measurements, the zero-thermal critical switching current $I_{c0}$, thermal stability $\Delta$, and effective damping-like SOT (DL-SOT) efficiency $\xi_{\text{DL}}^{\text{eff}}$ of these W-based heterostructures can be further determined. $|\xi_{\text{DL}}^{\text{eff}}| \approx 0.32$ is estimated for devices with $2.0 \text{ nm} \leq t_{\text{Co-Fe-B}} \leq 3.5$ nm, which is fairly consistent with the magnitude for thin W layer obtained by other approaches [1,16-18]. Field-like SOT (FL-SOT) efficiency can be simultaneously estimated and is found to be influenced by placing a MgO layer on top of Co$_{40}$Fe$_{40}$B$_{20}$, with a magnitude much smaller than its damping-like counterpart ($|\xi_{\text{FL}}^{\text{eff}}| \leq 0.05$). Lastly, we demonstrate that for heterostructures with a well-defined in-plane easy axis, as prepared from an 8-inch CMOS-compatible fabrication facility, the probing of SOT switching can also be realized by the UMR readouts. Our results suggest that this pump (DL-SOT)-probe (AMR or UMR) method can be used to determine key parameters of SOT switching phenomenon from IMA heterostructures without complicated fabrication and measurement processes.

## II. MATERIALS SYSTEMS

A series of W(3)/Co$_{40}$Fe$_{40}$B$_{20}$($t_{\text{Co-Fe-B}}$)/Ta(2) (numbers in parentheses are in nanometers) multilayer structures are deposited onto Si/SiO$_2$ substrate by high-vacuum magnetron sputtering (base pressure ~ 10$^{-8}$ Torr), with $t_{\text{Co-Fe-B}}$ ranges from 1.4 nm to 4.5 nm. A thin, amorphous, and resistive W(3) spin Hall material layer is chosen to ensure a sizable SHE can be produced. The Ta(2) will naturally oxidize in air (denoted as TaO$_x$ with $\rho_{\text{TaO}_x} \approx 850\ \mu\Omega \cdot \text{cm}$, which is way more resistive than other layers) and serves as a capping layer to prevent further oxidation of the underlayers [19] (See supplementary material S1 for TEM inspection of the deposited films). We characterize the in-plane anisotropy and the saturation magnetization of these as-deposited W/Co$_{40}$Fe$_{40}$B$_{20}$ films by vibrating sample magnetometer (VSM). The coercive fields for these films remain constants ($H_c \approx$ 20 Oe) by sweeping in-plane magnetic fields along various different directions, which suggests an easy-plane-like anisotropy for our samples. Saturation magnetization of the Co$_{40}$Fe$_{40}$B$_{20}$ layer for this series is found to be $M_s \approx 700\ \text{emu/cm}^3$, which is smaller than that from a standard Co$_{40}$Fe$_{40}$B$_{20}$. The reduction of $M_S$ might be originate from the unexpected larger B-concentration [20] and amorphous phase [19,21] for CoFeB layer after sputter deposition. For devices used in AMR and current-induced SOT switching measurements, we pattern the as-deposited films into micron-sized Hall-bar devices with channel width of 5 μm through standard photolithography and subsequent Ar-ion milling processes.

## III. SPIN HALL EFFECTIVE FIELD AND USMR

As schematically shown in Fig. 1(a), to measure magnetoresistance with four-point probe method, we sweep the in-plane magnetic field along $y$ direction ($H_y$) while applying a dc current along $x$ direction ($I_{dc}$). Representative normalized AMR loops for a W(3)/Co$_{40}$Fe$_{40}$B$_{20}$(1.8) sample with $I_{dc}=\pm 3$mA are shown in Fig. 1(b). Two features in these AMR loops can be identified: A slight horizontal shift in between two peaks of the AMR results (denoted as $H_{\text{eff}}^y$) and a vertical shift (denoted as $R_{H_-}^{I_+}-R_{H_-}^{I_-}$) as the magnitude of $H_y$ is large enough to saturate magnetization along $y$ direction. The horizontal shifts along $y$ direction in AMR loops with $I_{dc}=\pm 3$mA indicate the existence of a current-induced effective field $H_{\text{eff}}^y$, while the vertical shifts suggest the existence of an UMR that depends on both the direction of $I_{dc}$ and the orientation of magnetization $M$ with respect to $y$ axis. To investigate the origins of these phenomena, we measure the same AMR signals from a Pt-based sample, Pt(6)/Co$_{40}$Fe$_{40}$B$_{20}$(2.5), and find that both the trends of AMR loop shifts $H_{\text{eff}}^y$ (Fig. 1(c) and (d)) and $\Delta R_{\text{step}} \equiv [(R_{H_+}^{I_+}-R_{H_+}^{I_-})-(R_{H_-}^{I_+}-R_{H_-}^{I_-})]$ (Fig.1(e)) are reversed when the HM layer is changed from W to Pt. Since Pt and W possess opposite signs of the spin Hall ratio, our results therefore suggest that the current-induced $H_{\text{eff}}^y$ and $\Delta R_{\text{step}}$ are both having a SHE-related origin. We call this effective field the spin Hall field (SHF), which has also been observed in other HM/FM heterostructures when the FM layer is thin ($\leq 4$ nm) [22]. The UMR effect $\Delta R_{\text{step}}$, on the other hand, is most likely to be originated from the unidirectional spin Hall magnetoresistance (USMR) [23-27], which is caused by the variation between spin accumulation vector in HM/FM interface and magnetization of the FM layer. The magnitude of USMR is typically defined as $(R_{H_+}^{I_+}-R_{H_-}^{I_+})$, $(R_{H_-}^{I}-R_{H_+}^{I})$ or $\Delta R_{\text{step}}/2$. The magnitude of our USMR obtained by dc measurements in Pt and W are also in agreement with a previous report as obtained by an ac (2nd harmonic) approach [23].

The $Co_{40}Fe_{40}B_{20}$ thickness dependence of USMR/$I_{dc}$ and $H_{\text{eff}}^{y}/I_{dc}$ obtained from the W(3)/$Co_{40}Fe_{40}B_{20}(t_{\text{Co-Fe-B}})$ devices are summarized in Fig. 1(f). Both USMR/$I_{dc}$ and $H_{\text{eff}}^{y}/I_{dc}$ decrease with increasing $Co_{40}Fe_{40}B_{20}$ thickness, as expected, due to the reduction of spin current by reflection and current shunting in the bilayer structure [23,26,27]. The $Co_{40}Fe_{40}B_{20}$ thickness dependence of USMR/$I_{dc}$ and $H_{\text{eff}}^{y}/I_{dc}$ are both proportional to $1/(t_{\text{Co-Fe-B}}-t_c)$, where the critical thickness for FM layer $t_c \approx 1.3$ nm. For films with $t_{\text{Co-Fe-B}} < t_c$, PMA starts to emerge. Note that $H_{\text{eff}}^{y}/I_{dc}$ can reach as large as -24 Oe/mA in the W(3)/$Co_{40}Fe_{40}B_{20}(1.4)$ sample, which is much greater than the expected Oersted field from such layer structure. The direction of the observed $H_{\text{eff}}^{y}$ is also opposite to that of the Oersted field. Also note that it is possible to have thermal contribution in the detected USMR signal [23,24,26]. However, this thermal effect will not affect the SOT switching measurement and characterization protocol that will be discussed in the following sections.

## IV. SPIN-ORBIT TORQUE SWITCHING

Next, we demonstrate that current-induced SOT switching of a FM layer with IMA can be probed by AMR measurements with the aid of the SHF. The protocol of the measurement is straightforward: We apply a write current pulse $I_{\text{write}}$ along $x$ to switch the magnetization $M$ between $\pm y$ directions with DL-SOT and inject a sense current $I_{\text{sense}}$ to detect the SHF-modified AMR signal. With the measurement sequences shown in Fig. 2(a) and (c), we measure the longitudinal resistance $R_{xx}$ by sweeping the write current pulses from negative to positive values and back to negative (details of pulsed current can be found in supplementary material S2). It is found that $R_{xx}$ of a representative W(3)/Co$_{40}$Fe$_{40}$B$_{20}$(1.8) device can be switched between a high resistance state and a low resistance state at critical switching currents of $I_c \approx \pm 1.2$ mA with $I_{\text{sense}} = +1$ mA (Fig. 2(b)) or $I_{\text{sense}} = -1$ mA (Fig. 2(d)). The measured critical switching current is reproducible even after 100 switching cycles (supplementary material S3) and is not strongly affected by the amplitude of sense currents (supplementary material S4). Note that the polarities of the current-induced switching loops are opposite for using sensing currents of opposite signs.

The origin of these two different resistance states in $R_{xx}$ is schematically shown in Fig. 2(e). In the absence of sense current ($I_{\text{sense}} = 0$ mA), the ideal AMR response is not shifted due to the absence of SHF. Under this circumstance, $R_{xx}$ is not distinguishable between the states with $\boldsymbol{M}$ pointing along $\pm y$ directions. When $I_{\text{sense}}$ is non-zero, the states with $\boldsymbol{M}$ pointing along $+y$ and $-y$ directions will have different $R_{xx}$ values due to the influence from the SHF. The corresponding resistance for $\boldsymbol{M}$ pointing along $\pm y$, either being high or low, will also depend on the sign of $I_{\text{sense}}$. Therefore, this SHF-modified AMR can be used to probe the direction of $\boldsymbol{M}$. When $I_{\text{write}}$ is applied, the SHE generated in the W layer will also result in an in-plane (anti) damping-like torque [28,29] acting on the Co$_{40}$Fe$_{40}$B$_{20}$ layer. When the write current is large enough, this DL-SOT will overcome the intrinsic damping and switch $M$ along $y$ axis, which can be further detected by the variation of the SHF-modified AMR.

To further enhance the detected signal, we record the difference of resistance $\Delta R = \left(R_{xx}^{I+} - R_{xx}^{I-}\right)$ using opposite sense currents ($I_{\text{sense}} = \pm 1$ mA) during field sweep (Fig. 3(a)) and write current sweep (Fig. 3(b)). Representative field sweep and current sweep results are respectively shown in Fig. 3(c) and (d). The presented raw data show that both field-driven and current-driven switching will cause a reversible variation of $\Delta R$ between -4.0 Ω and -7.5 Ω with $I_{\text{sense}} = \pm 1$ mA, which suggests an almost full magnetization switching caused by the DL-SOT under zero-field condition. We further perform current-induced switching measurements with different pulse widths (0.05 s ≤ $t_{\text{pulse}}$ ≤ 1 s) of $I_{\text{write}}$. Since SOT-driven magnetization switching is a thermally-activated process, the dependence of critical switching current $I_c$ on write current pulse width $t_{\text{pulse}}$ can be expressed as [28]:

$$I_c = I_{c0}\left[1 - \frac{1}{\Delta}\ln\left(\frac{t_{\text{pulse}}}{\tau_0}\right)\right], \tag{1}$$

where $I_{c0}$ is the zero-thermal-fluctuation critical switching current, $\Delta \equiv U/k_B T$ is the thermal stability factor ($U$ being the energy barrier between two magnetic states), and $\tau_0 \approx 1$ ns is the attempt rate for thermally-activated switching [30]. By linearly fitting the pulse width dependence switching results, as shown in Fig. 3(e), we find that $I_{c0} \approx \pm 2.3$ mA ($J_{c0} \approx \pm 9.57 \times 10^{10}$ A/m²) and $\Delta \approx 53$ for the W(3)/Co$_{40}$Fe$_{40}$B$_{20}$(1.8) Hall-bar device. The spin-torque switching efficiency $\varepsilon \equiv \Delta/I_{c0}$ [31], which is considered as the key figure of merit for addressing STT-MRAM performance, is further quantified for all W(3)/Co$_{40}$Fe$_{40}$B$_{20}$($t_{\text{Co-Fe-B}}$) Hall-bar devices. As shown in Fig. 3(f), the value of $\varepsilon$ is estimated to be $\sim 2\times10^{-2}$ µA$^{-1}$ for these 5-µm-wide devices.

Once the zero-thermal-fluctuation critical current density $J_{c0}$ is obtained from switching data, the DL-SOT efficiency of a heterostructure with IMA can be calculated by [7,28,32] :

$$\xi_{\text{DL}}^{\text{eff}} = \frac{2e\mu_0 M_s t_{\text{FM}}}{\hbar}\left[\frac{\alpha_0\left(H_c+\frac{M_{\text{eff}}}{2}\right)}{J_{c0}}\right], \qquad (2)$$

and the FL-SOT efficiency can be obtained by [33,34]:

$$\xi_{\text{FL}}^{\text{eff}} = \frac{2e\mu_0 M_s t_{\text{FM}}}{\hbar}\left(\frac{H_T}{J_e}\right), \qquad (3)$$

where $M_s$, $\mu_0 M_{\text{eff}}$, and $\alpha_0$ are saturation magnetization, effective demagnetization field, and damping constant of the Co$_{40}$Fe$_{40}$B$_{20}$ layer, respectively. $H_T = H_{\text{eff}}^y - H_{\text{Oe}}$ is the net SHF (excluding the Oersted field), which is proportional to the current density $J_e$ applied in the W layer. $\alpha_0 \approx 0.02$ and $\mu_0 M_{\text{eff}} \approx 1.1$ T are further determined by spin torque-ferromagnetic resonance (ST-FMR) measurement [4]. Using these parameters with Eqn. (2) and Eqn. (3), we can estimate the effective DL-SOT efficiencies (by current-induced switching measurements) and FL-SOT efficiencies (by AMR loops shift measurement) for W(3)/Co$_{40}$Fe$_{40}$B$_{20}$($t_{\text{Co-Fe-B}}$) heterostructures. As shown in Fig. 4(a), the DL-SOT efficiency decreases from $|\xi_{\text{DL}}^{\text{eff}}| \approx 0.58$ for $t_{\text{Co-Fe-B}} = 1.8$ nm to $|\xi_{\text{DL}}^{\text{eff}}|_{\text{avg}} \approx 0.32$ for $2.0 \text{ nm} \leq t_{\text{Co-Fe-B}} \leq 3.5 \text{ nm}$, which is fairly consistent with the values as obtained by TMR measurement ($|\xi_{\text{DL}}^{\text{eff}}| \approx 0.33$) [1], MOKE measurement ($|\xi_{\text{DL}}^{\text{eff}}| \approx 0.40$) [17] and inverse spin Hall effect (ISHE) measurement ($|\xi_{\text{DL}}^{\text{eff}}| \approx 0.44$) [18]. The slight increase of $|\xi_{\text{DL}}^{\text{eff}}|$ from $t_{\text{Co-Fe-B}} = 3$ nm to 3.5 nm is following the trend of ST-FMR-determined damping constant, which suggests that not only the critical switching current density $J_{c0}$, but also the damping constant $\alpha_0$ of the measured device will affect the estimation of DL-SOT efficiency. To compare, the FL-SOT efficiency is much smaller than its DL counterpart and decreases from $|\xi_{\text{FL}}^{\text{eff}}| \approx 0.054$ for $t_{\text{Co-Fe-B}} = 1.8$ nm to $|\xi_{\text{FL}}^{\text{eff}}| \approx 0.014$ for $t_{\text{Co-Fe-B}} = 3.5$ nm.

To examine the effect of oxide capping layer to the DL and FL-SOT efficiencies, we also prepare and test on a series of W(3)/Co$_{40}$Fe$_{40}$B$_{20}$(2.5)/MgO($t_{\text{MgO}}$)/TaO$_x$ samples, with 0 nm $\leq t_{\text{MgO}} \leq$ 2 nm. We find that as $t_{\text{MgO}}$ increases from 0 nm to 1.2 nm, $|\xi_{\text{FL}}^{\text{eff}}|$ also gradually increases and saturates at $|\xi_{\text{FL}}^{\text{eff}}| \approx 0.05$. In contrast, the DL-SOT efficiency for this series of samples ($|\xi_{\text{DL}}^{\text{eff}}| \approx 0.30$) does not show significant dependence on $t_{\text{MgO}}$. The slight enhancement of FL-SOT efficiency might be originated from the elimination of the SHF from Ta layer or the enhancement of in-plane effective field from the Rashba effect at the Co$_{40}$Fe$_{40}$B$_{20}$/MgO interface [8,35]. However, it is shown that the MgO layer has little effect on the obtained DL-SOT efficiency.

To further apply this approach, a layer stack of W(4)/Co$_{40}$Fe$_{40}$B$_{20}$(1.4)/MgO(2.1)/Ta(10) is prepared and made into the same micron-sized Hall-bar devices by 8-inch CMOS-compatible fabrication processes. The devices are further annealed at 360 degree C for 20 minutes with applying an in-plane magnetic field of $10^4$ Oe to gain easy axis (EA). The anisotropy constant $K_u$= 3.9×$10^4$ J/$m^3$ is further determined by VSM measurement. The annealing of $Co_{40}Fe_{40}B_{20}$ layer at this elevated temperature with large external field can promote boron diffusion to enhance the saturation magnetization of $Co_{40}Fe_{40}B_{20}$ and further induce magnetocrystalline anisotropy (MCA) along a specific direction to gain EA. Two types of Hall-bar devices are patterned: One with the current channel parallel to the EA and the other with the current channel perpendicular to the EA. Fig. 5(a) and (b) show representative field-sweep AMR loops ($H \perp I$) as obtained by applying current along the EA ($I \parallel$EA) and perpendicular to the EA ($I \perp$EA), respectively. A hard axis behavior can be observed for the $I \parallel$EA sample, whereas the $I \perp$EA sample shows clear hysteresis loop behavior with two distinct resistance states, which is attributed to the USMR. Note that a similar switching signal has been seen in an epitaxial paramagnet-(Ga,Mn)As/ferromagnet-(Ga,Mn)As bilayer system [36], but the switching mechanism therein is mainly attributed to the Oersted field. Also note that some smaller hysteresis jumps can be seen in Fig. 5(a) due to the imperfect alignment of the applied field with respect to the hard axis. $\Delta R$ of the device with $I \perp$EA is further recorded by sweeping magnetic field (Fig. 5(c)) and pulsed currents (Fig. 5(d)). Note that the intermediate states in Fig. 5(c) are caused by the SHF. Full current-induced magnetization switching via DL-SOT can be observed with $I_c \approx \pm 10$ mA ($J_c \approx \pm 1.99 \times 10^{11}$ A/$m^2$, assuming that the oxidized thickness in Ta is 2 nm). For this particular device with a well-defined EA, the effect of SHF-modified AMR may not be necessary for detecting current-induced switching due to the presence of two obvious resistance states from USMR. It suggests that our approach can be further simplified if the films with IMA possess EA along *y* direction.

## V. CONCLUSION

To conclude, we first demonstrate a SOT-induced magnetization switching detection scheme via the SHF-modified AMR in micron-sized W/Co$_{40}$Fe$_{40}$B$_{20}$ devices, where the Co$_{40}$Fe$_{40}$B$_{20}$ layer is in-plane magnetized. Through this method, we can estimate both the effective DL-SOT efficiency $\xi_{\mathrm{DL}}^{\mathrm{eff}}$ and the FL-SOT efficiency $\xi_{\mathrm{FL}}^{\mathrm{eff}}$ of various W/Co$_{40}$Fe$_{40}$B$_{20}$ heterostructures. Other key parameters for SOT switching phenomenon such as thermal stability $\Delta$ of the FM layer and the switching efficiency ($\varepsilon \equiv \Delta / I_{\mathrm{c0}}$) can also be quantified from the switching data. For a representative micron-sized device that has a well-defined easy axis as fabricated from an 8-inch CMOS fab facility, the SOT-driven switching can also be probed by USMR measurement. The development of this measurement protocol therefore allows for mitigating the complexness of characterizing SOT-driven switching parameters in micron-sized magnetic heterostructures with IMA.

**Acknowledgments**

This work is supported by the Ministry of Science and Technology of Taiwan (MOST) under grant No. MOST 108-2636-M-002-010 and by the Center of Atomic Initiative for New Materials (AI-Mat),

National Taiwan University, Taipei, Taiwan from the Featured Areas Research Center Program within the framework of the Higher Education Sprout Project by the Ministry of Education (MOE) in Taiwan under grant No. NTU-107L9008. We also acknowledge the financial and technical supports from the Industrial Technology Research Institute (ITRI) in Taiwan. We thank Tsao-Chi Chuang for her assistance on VSM measurements, Chun-Hao Chiang and Chen-Yu Hu for his assistance on AFM measurements, Cheng-Yu Chen for his assistance on TEM imaging and Chia-Ying Chien of Ministry of Science and Technology (National Taiwan University) for the FIB experiments.

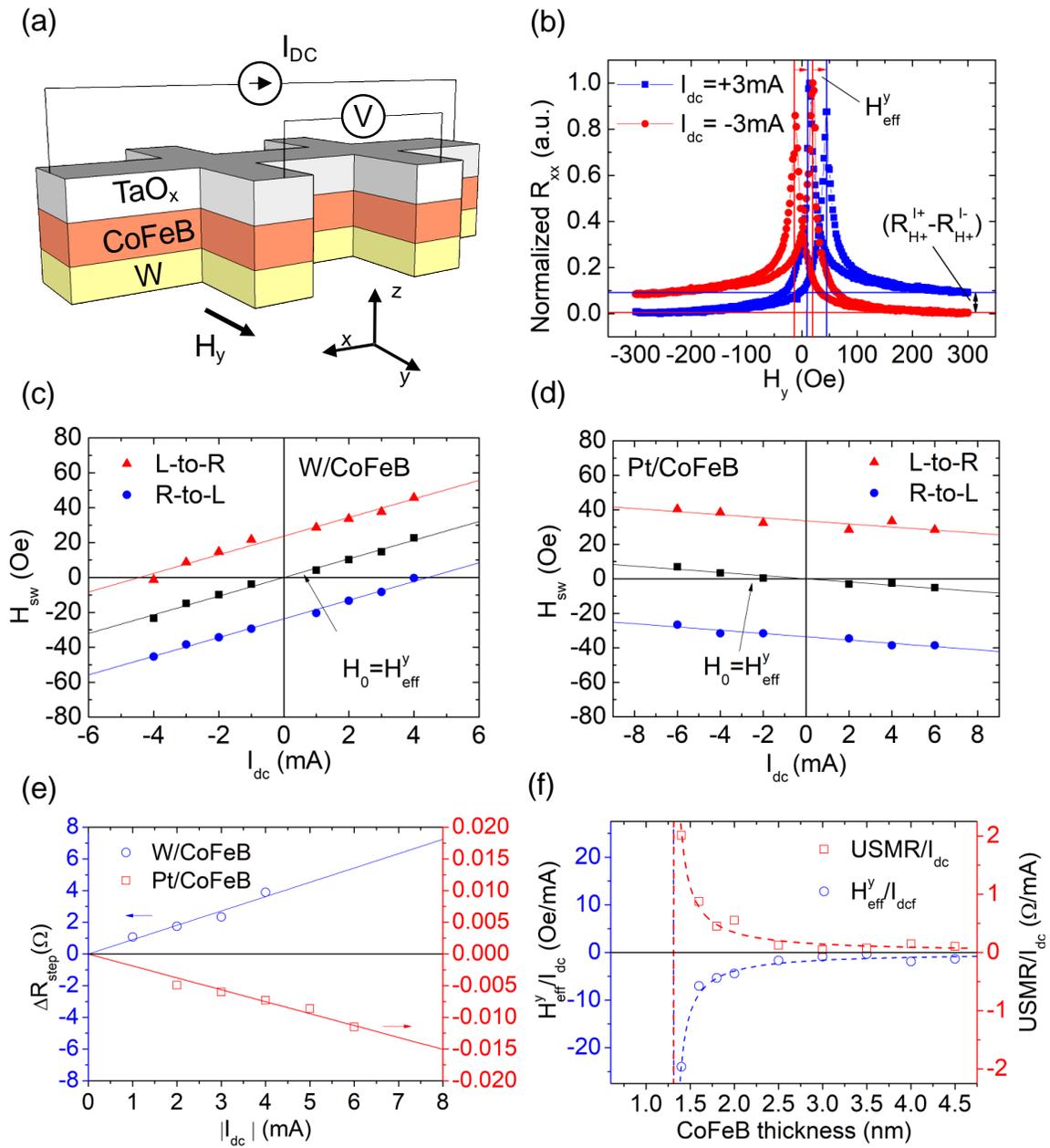

Figure 1. (a) Schematic illustration of W/Co-Fe-B Hall-bar device with lateral dimensions of 5 μm × 60 μm and AMR measurement. The direction of the positive current defined along the -*x* direction. (b) Representative shifted AMR loops obtained from a W(3)/Co$_{40}$Fe$_{40}$B$_{20}$(2.0) sample with dc current $I_{dc}$ = ± 3mA. Switching field $H_{sw}$ of (c) W(3)/Co$_{40}$Fe$_{40}$B$_{20}$(1.8) and (d) Pt(6)/Co$_{40}$Fe$_{40}$B$_{20}$(2.5) sample

for left(-y)-to-right(+y) (L-to-R) and right(+y)-to-left(-y) (R-to-L) switching processes as functions of $I_{dc}$. $H_{eff}^{y}$ represents the center of the AMR loop peaks. (e) $\Delta R_{step}$ versus $I_{dc}$ for W(3)/Co$_{40}$Fe$_{40}$B$_{20}$(1.8) and Pt(6)/Co$_{40}$Fe$_{40}$B$_{20}$(2.5) samples. (f) $H_{eff}^{y}/I_{dc}$ and USMR/$I_{dc}$ of W(3)/Co$_{40}$Fe$_{40}$B$_{20}$($t_{Co-Fe-B}$) samples as functions of Co$_{40}$Fe$_{40}$B$_{20}$ thickness.

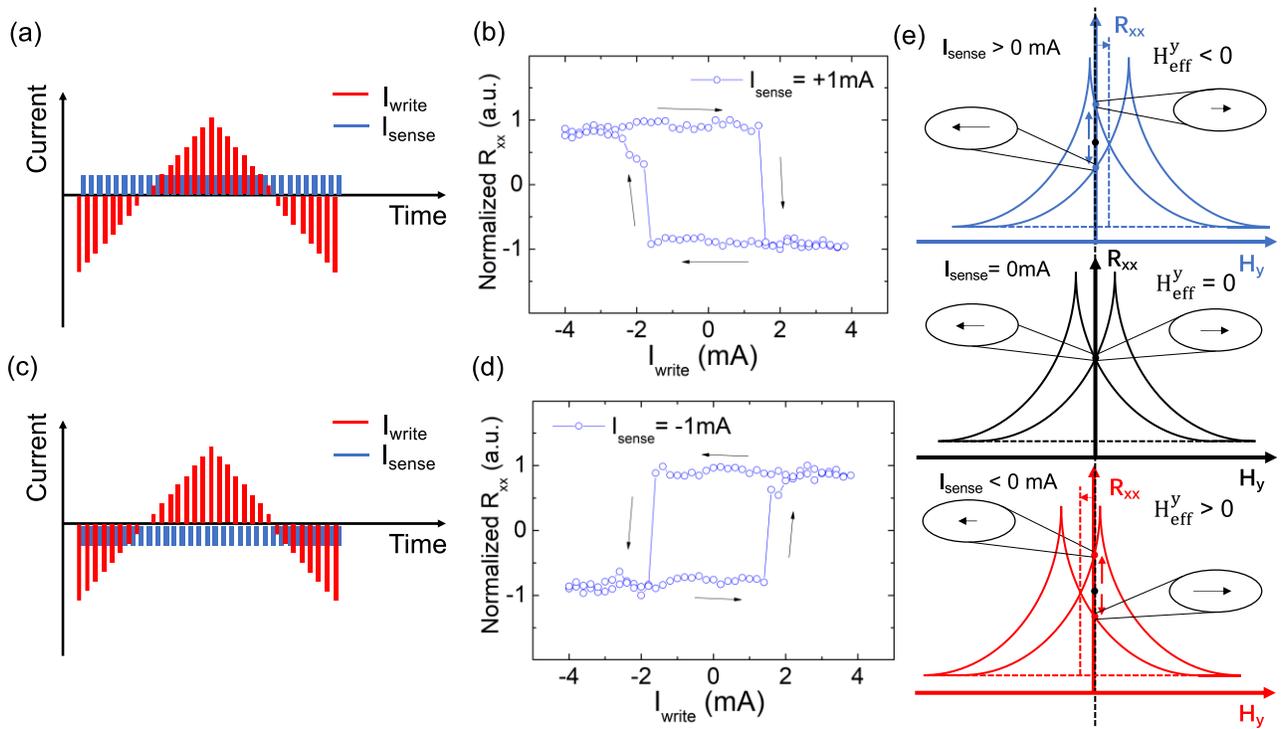

Figure 2. (a, c) Measurement sequence for the switching experiment with opposite sense currents. (b, d) Current-induced DL-SOT switching results of a W(3)/Co$_{40}$Fe$_{40}$B$_{20}$(1.8) Hall-bar sample with opposite sense currents $I_{sense}$ = + 1 mA and $I_{sense}$ = -1 mA. The black arrows indicate the switching direction. (e) Schematics of current-induced AMR loops shift (SHF-modified AMR). The perpendicular dashed line represents the center of two peaks of AMR by applying different sense currents.

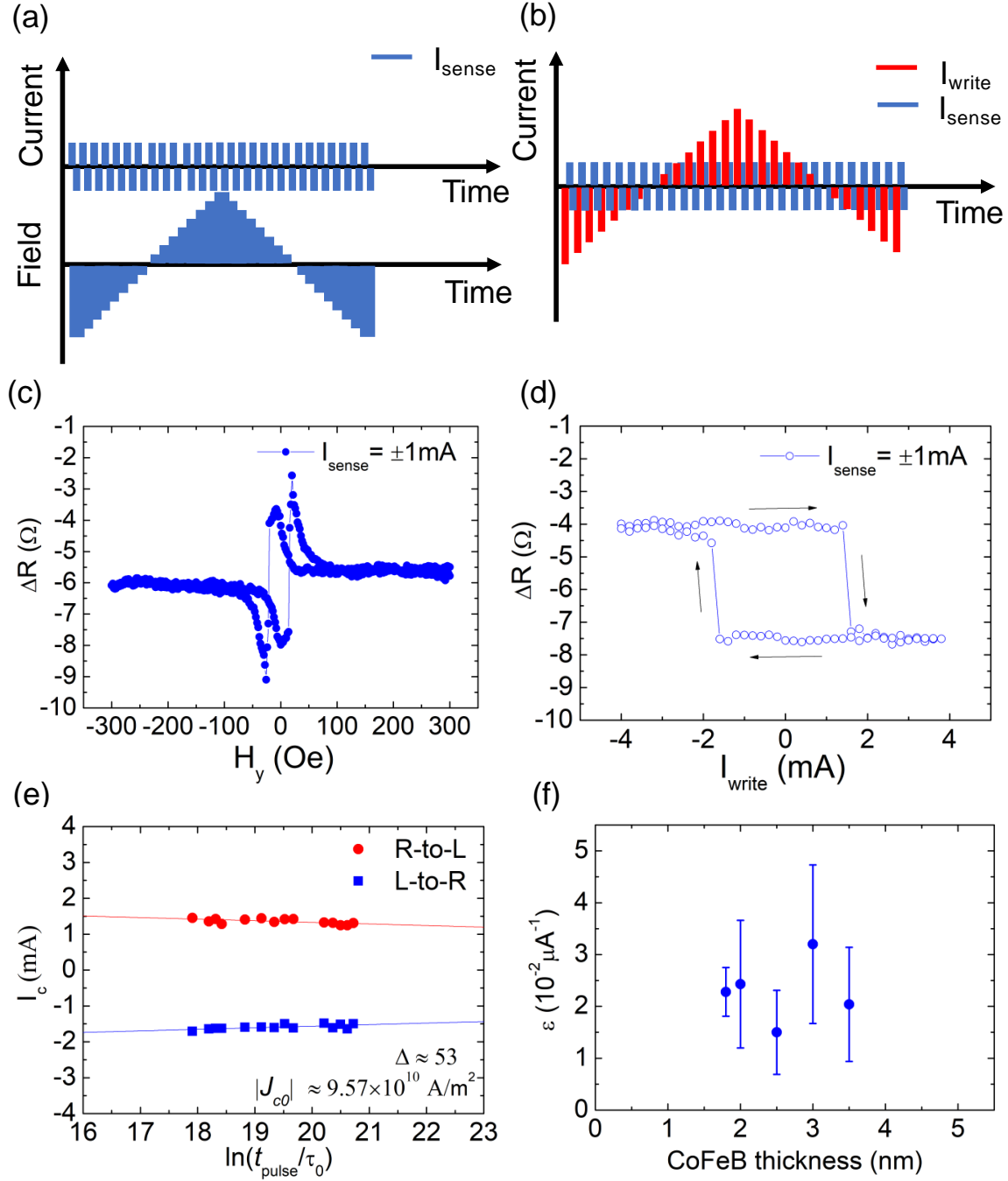

Figure 3. Measurement sequence of $\Delta R$ by sweeping (a) in-plane magnetic field $H_y$ and (b) write current $I_{write}$. Representative $\Delta R$ as a function of (c) in-plane magnetic field $H_y$ and (d) write current $I_{write}$ for a W(3)/Co$_{40}$Fe$_{40}$B$_{20}$(1.8) device. (e) The write current pulse width dependence of critical switching current $I_c$ for a W(3)/Co$_{40}$Fe$_{40}$B$_{20}$(1.8) device. The solid lines represent linear fits to the experimental data. (f) Spin-torque switching efficiency of W(3)/Co$_{40}$Fe$_{40}$B$_{20}$($t_{\text{Co-Fe-B}}$) samples as function of Co$_{40}$Fe$_{40}$B$_{20}$ thickness.

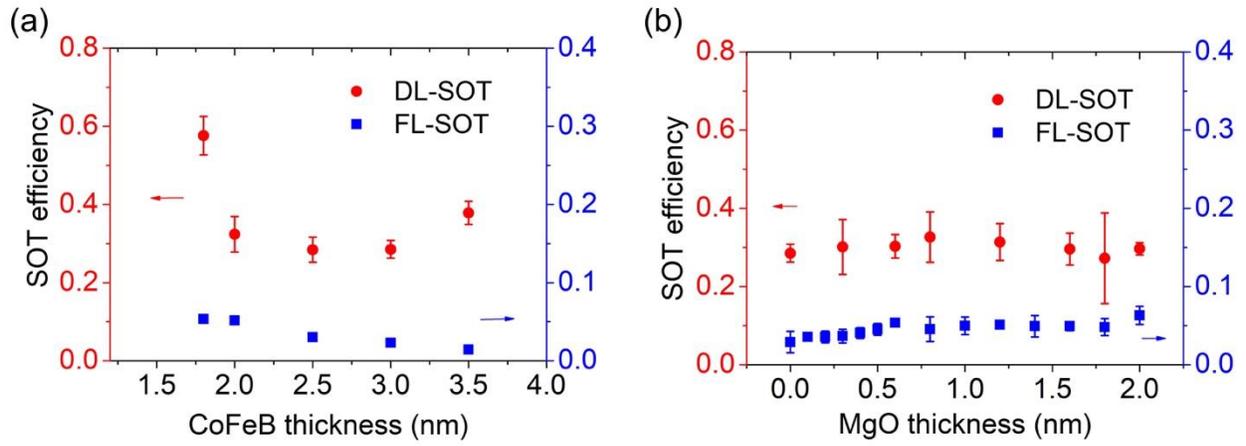

Figure 4. (a) The DL-SOT efficiency and FL-SOT efficiency of W(3)/Co$_{40}$Fe$_{40}$B$_{20}$($t_{\text{Co-Fe-B}}$) samples as functions of Co$_{40}$Fe$_{40}$B$_{20}$ thickness. (b) The DL-SOT efficiency and FL-SOT efficiency of W(3)/Co$_{40}$Fe$_{40}$B$_{20}$(3.0)/MgO($t_{\text{MgO}}$) samples as functions of MgO thickness. Some uncertainties are smaller than the symbol size shown.

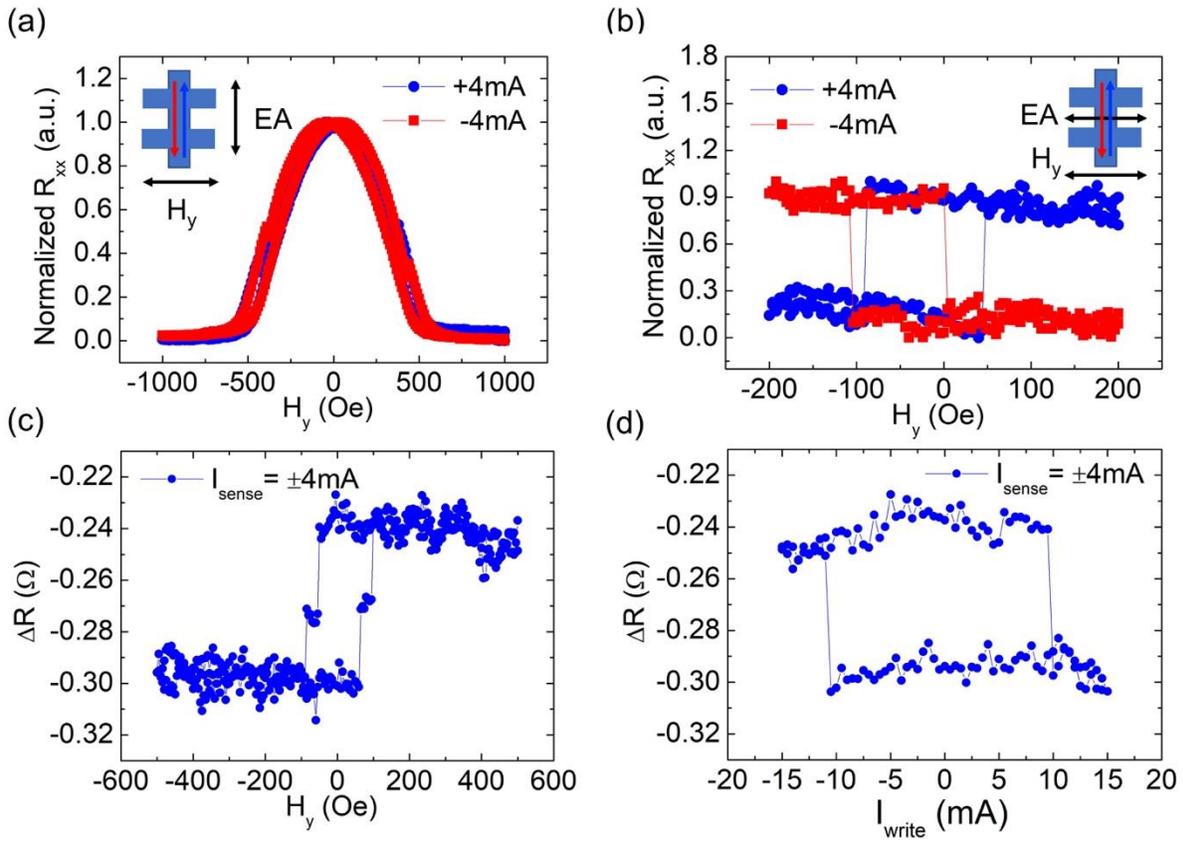

Figure 5. Normalized AMR results from W(4)/Co$_{40}$Fe$_{40}$B$_{20}$(1.4)/MgO(2.1)/Ta(10) devices by applying currents $I_{sense}$ = ± 4 mA (a) parallel to the easy axis ($I$∥EA) and (b) perpendicular to the easy axis ($I$⊥EA). $\Delta R$ as functions of (c) in-plane magnetic field $H_y$ and (d) write current $I_{write}$ for a W(4)/Co$_{40}$Fe$_{40}$B$_{20}$(1.4)/MgO(2.1)/Ta(10) device with $I$⊥EA.